# Tailoring Luminescent Properties of SrS:Ce by Modulating Defects: Sr-Deficiency and Na$^+$ Doping

*Shuqin Chang[†], Jipeng Fu[†,‡,\*], Xuan Sun[†,‡], Guangcan Bai[§], Guoquan Liu[§], Kaina Wang[‡], Ligang Xu[Δ], Qi Wei[Δ], Thomas Meier[†], Mingxue Tang[†,\*]*

[†]Center for High Pressure Science and Technology Advanced Research, 100094 Beijing, China

[‡]Key Laboratory of Rare Earth Optoelectronic Materials and Devices of Zhejiang Province, Institute of Optoelectronic Materials and Devices, 258 Xueyuan street, China Jiliang University, 310018 Hangzhou, China

[§]State Key Laboratory of Natural and Biomimetic Drugs, School of Pharmaceutical Sciences, Peking University, 100191Beijing, China

[Δ]College of Materials Science and Engineering, Beijing University of Technology, No. 100 Pingleyuan, 100124 Beijing, China

Jipeng.Fu@hpstar.ac.cn

Mingxue.tang@hpstar.ac.cn

Ce$^{3+}$ doped SrS phosphors with a charge-compensating Na$^+$ for light-emitting diode (LED) applications have been successfully synthesized via a solid-state reaction method, which can be indexed to rock-salt-like crystal structures of Fm-3m space group. SrS:(Ce$^{3+}$)$_x$ (0.005≤x≤0.05) and SrS:(Ce$^{3+}$)$_{0.01}$,(Na$^+$)$_y$ (0.005≤y≤0.030) phosphors were excited by 430 nm UV-VIS light, associated to the 5d$^1$ → 4f$^1$ transition of Ce$^{3+}$. The composition-optimized SrS: (Ce$^{3+}$)$_{0.01}$, ($Na^+$)$_{0.015}$ phosphors showed an intense broad emission band at λ=430-700nm. The doping of $Na^+$ was probed by solid-state nuclear magnetic resonance. The 430 nm pumped white LED (w-LED) combining SrS:($Ce^{3+}$)$_{0.01}$,($Na^+$)$_{0.015}$ phosphors and $Sr_2Si_5N_8$:$Eu^{2+}$ phosphors shows a color-rendering index (Ra) of 89.7. The proposed strategy provides new avenues for design and realization of novel high color quality solid-state lighting emitting diodes (SS-LEDs).

**TOC GRAPHICS**

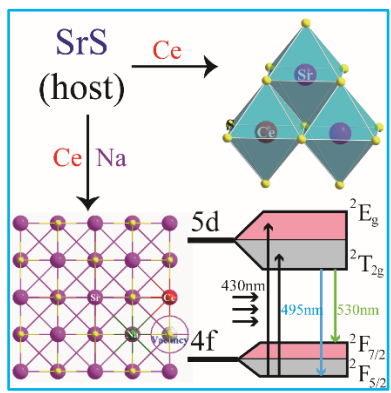

Phosphor-converted white light-emitting diode (pc-wLED) lighting has acquired incredible accomplishment since the successful invention of the InGaN blue light-emitting diode by Isamu Akasaki et al., awarded the Nobel Physics Prize in 2014.[1] Currently, pc-wLEDs have become an increasingly hot topic due to a wealth of unique advantages, such as environmental friendliness, long life-time, and tunable colors allowing for a wide utilization in the fields of bio-imaging, anti-counterfeiting, lighting, medical applications and many more.[2–6] Exploring classical phosphors, being the essential ingredient of SS-LEDs, has been performed by introducing rare earth (RE) ions into inorganic host materials.[7–9] In the early 1990s, the phosphors of alkali-earth metal sulfides (MgS, CaS, SrS and BaS) doped with $Eu^{2+}$, $Eu^{3+}$ and $Ce^{3+}$ were extensively studied for cathodoluminescence and electroluminescence displays.[10,11] A series of commercial CaS:Eu phosphors have recently been investigated as solid state light sources due to the red emission near 645 nm. SrS is considered as a potential substrate matrix owing to its small vibrational frequencies and narrow bandgap energy.[12,13] The luminescence properties of SrS can be modulated over the entire region of the visible electro-magnetic spectrum when doped with appropriate RE ions. However, conventional synthesis methods of alkaline-earth sulfide compositions including gas-solid diffusion reactions, reverse micelles, metal organic chemical vapor deposition and solvothermal methods, restricted the large-scale preparation of strontium sulfide and application in white light LEDs.[14-16] SrS phosphors were obtained via gas-solid reaction method by sulfurating alkaline earth-based sulfate, carbonate and nitrate in $N_2$ with a $H_2S$ or $CS_2$ atmosphere.[17] This approach involved toxic gases and thus is prohibited for wider academic investigations and potential industrial applications. Therefore, it is of great importance to find alternative synthesis methods with low cost, easy manipulating and environmental friendliness.

Additionally, cerium-doped phosphors have gained increased interest due to the excellent photoluminescent properties and their compatibility with different host materials. The $5d^1 \rightarrow 4f^1$ transition of the $Ce^{3+}$ activator is sensitive to the crystal structure and the site symmetry of the charge environment at the coordination position. $Ce^{3+}$-doped samples normally show $5d^1 \rightarrow 4f^1$ emission in the ultraviolet but in the case of high crystal-field splitting, such as in garnets ($Gd_3Al_5O_{12}$), visible emission is observed.[18] The emission spectra of $Ce^{3+}$ of garnet series, $Mg_3(Y_{1-y}Gd_y)_2(Ge_{1-z}Si_z)_3O_{12}:Ce^{3+}$ (y=0-1,x=0,1), can be modified with variations in the Ce-O bond lengths arising from different dopant concentration and chemical substitution.[19] Alkali metals were used to dope the host matrix with the goal of modulating the $Ce^{3+}$ spectra. However, the doping sites and local environments could not to be investigated directly due to a lack of experimental probes until recently. Solid-state magnetic resonance has been proved to be a highly versatile tool to probe the local electronic environment, ground state electronic structure, defects and chemical surroundings providing a powerful alternative to other characterization methods.[20-25]

In this work, $Ce^{3+}$ and $Na^+$ are successfully incorporated into SrS, exhibiting broad-band emissions in the wavelength range of λ=430–700 nm. The luminescent properties for the target materials were tailored by the doping of $Ce^{3+}$ and $Na^+$, which can be easily adjusted to optimize color rendering index and chromaticity. XRD and NMR were carried out to investigate the local structure-property correlations in the host materials. Afterglow phenomena were observed originating in vacancy defects in the crystalline lattice. SrS:$Ce^{3+}$ and SrS:$Ce^{3+}$,$Na^+$ phosphors under λ=430 nm light excitation for w-LED were fabricated and shown tunable luminescent properties by introduction of dopant and defect-involved emissions into single-dopant activated phosphors.

The micrometer-sized (μm) phosphors of SrS: $(Ce^{3+})_x$ (x = 0.5, 1.0, 2.0, 3.0, 4.0, 5.0 mol%) with a nominal composition were synthesized via high temperature solid-state reaction methods.[26,27] Powder X-ray diffraction (PXRD) patterns of the product series are shown in Fig. 1a. All major diffraction peaks are attributed to a rock-salt-like structure of SrS and can be well indexed with an Fm-3m space group, consistent with the standard diffraction patterns of this material class (PDF card No.08-0489). No extra phases were found, indicating that the doping of $Ce^{3+}$ in the materials do not affect the SrS long range crystalline order. The peaks between 2θ=50° and 53° degrees show a small right shift for the $Ce^{3+}$ doped SrS when compared with the pure material (Fig. S1). We concur that this might be caused by lattice expansion due to $Ce^{3+}$ incorporation. To further understand the phase structure and site occupation of $Ce^{3+}$ ions, Rietveld refinement was performed for the sample with x= 0.01 $Ce^{3+}$ doped (SrS:1.0 mol% $Ce^{3+}$, as shown in Fig. 1b). The unit cell volume varies non-linearly with increasing rare earth element doping concentration (Fig. S1), suggesting that $Ce^{3+}$ ion could partially occupy atomic sites of $Sr^{2+}$ ion.[28-30] Reliability factors including Bragg R-pattern factor (Rp=2.488%), weighted-profile R factor (Rwp=1.362%) and goodness-of-fit parameter ($\chi^2$ =1.27) evidence the excellent quality and the dependability of the refinement's data. The refined cell parameters are a=b=c=6.016 Å, $\alpha=\beta=\gamma=90^o$ and V=217.73 Å$^3$, confirming a cubic rock-salt structure of the as-prepared particle phosphors. The cell parameters are slightly smaller than that of standard values. The shrinking of the crystal structure with $Ce^{3+}$ doped could be attributed to two reasons: (1) a mass of $Ce^{3+}$ ions (1.01Å, CN=6) occupy the bigger $Sr^{2+}$ sites (1.18Å, CN=6). (2) The charge difference between $Ce^{3+}$ and $Sr^{2+}$ would generate the cation vacancies in the process of substitution, which can enhance the degree of crystal lattice shrinkage (Fig.S1).[31]

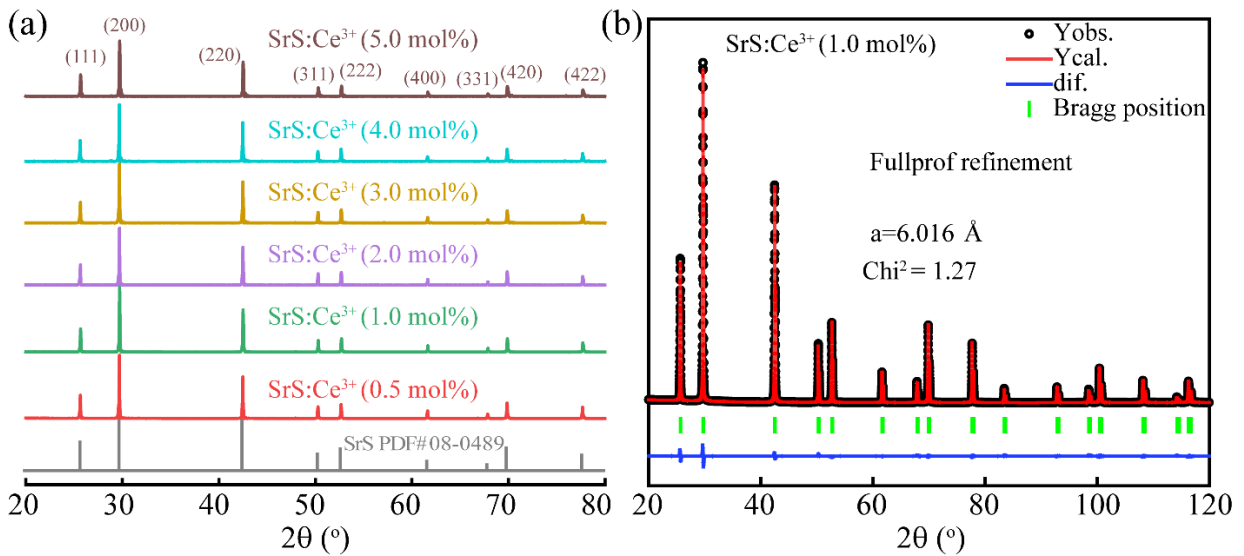

Figure 1. (a) X-Ray diffractograms of SrS with different doping content of $Ce^{3+}$. (b) Rietveld refinement of the SrS with 0.01 $Ce^{3+}$ dopant.

In order to characterize the morphology and elemental composition of the synthesized SrS:$(Ce^{3+})_{0.01}$ phosphor, field emission scan electron microscope (FE-SEM) and energy dispersive X-ray spectroscopy (EDS) were performed, Fig. S2. The SEM images show a homogenous distribution with an average particle size of 50 μm. The smooth surface would be beneficial to the luminescent properties and incorporation into LED packages. The elemental mapping images conducted on a randomly selected site show a uniform distribution of Sr, S, and Ce within the phosphor particles and the average atomic ratio of Sr:S is close to unity (Fig. S3), which is in good agreement with chemically anticipated stochiometry.

X-ray photoelectron spectroscopy (XPS) analysis was performed to investigate the chemical nature of the surface condition and get further insight into the electron effect of the doped phosphor host matrix. Fig. 2a shows XPS spectrum of SrS:$(Ce^{3+})_{0.01}$ in the binding energy range of E=0-900 eV, which indicates the presence of Sr, S, C, O in the system. Due to the low concentration of $Ce^{3+}$ dopants, a corresponding cerium signal was below our detection limits. The C-1s peak (E = 284.8 eV) corresponds to C-C, caused by adventitious carbon. The O-1s signal (E≈550eV) is likely caused by carbon dioxide absorption from air. High-resolution XPS spectra for $S^{2-}$ and $Sr^{2+}$ are shown in Fig 2b and 2c, respectively. S-2$p_{3/2}$ shows four peaks at E=168.6 eV, 166.7 eV, 162.1 eV, 160.1 eV.[32] Careful analysis of Sr-3d shows one doublet located at E=135.1 eV and 133.38 eV with an energy difference of 1.7 eV, associated to Sr $3d_{3/2}$ and Sr $3d_{5/2}$, respectively.[33]

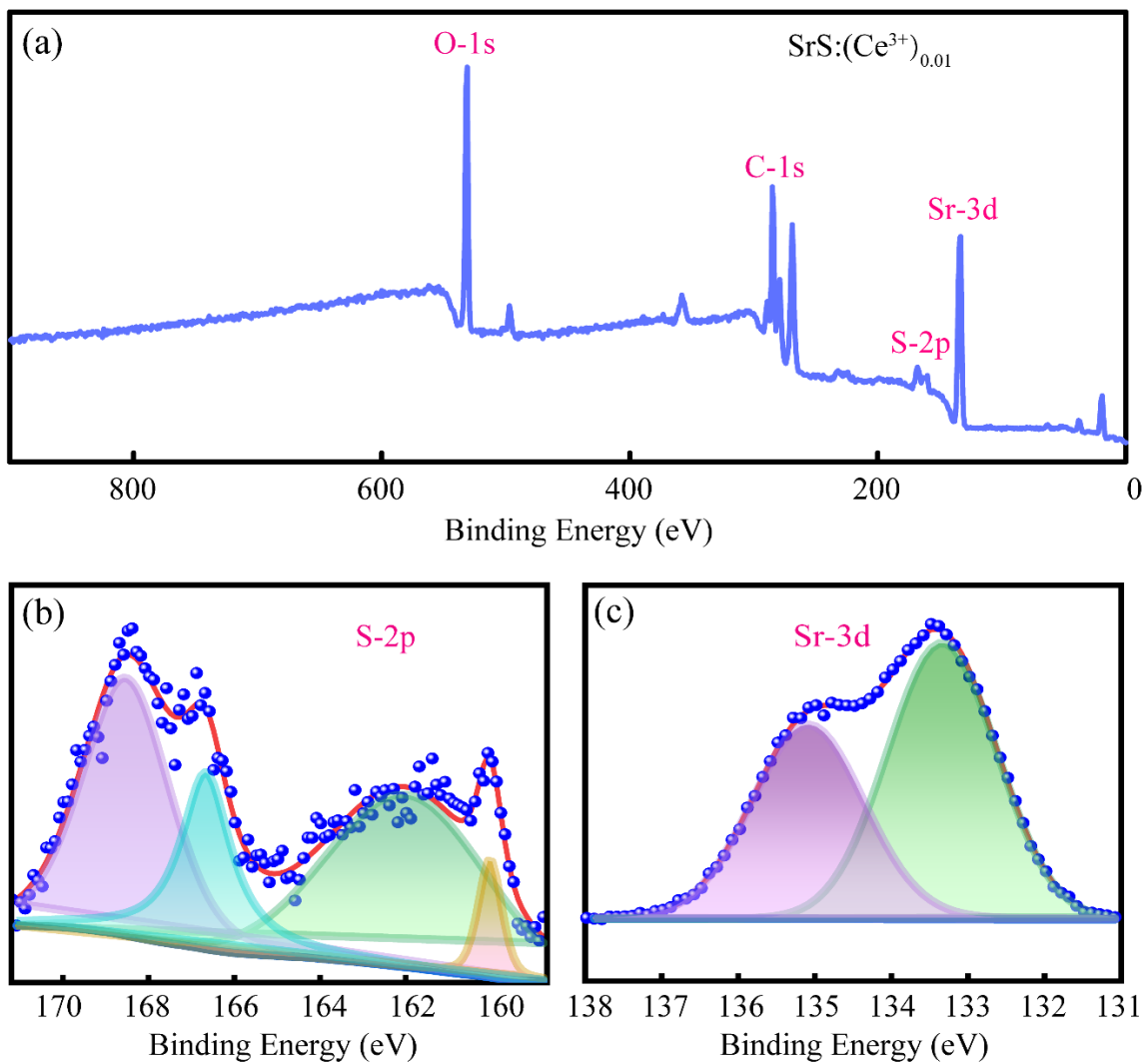

Figure 2. (a) Full XPS spectrum of SrS with x=0.01 $Ce^{3+}$ dopant. (b) High-resolution spectra of S-2p and (c) Sr-3d.

The photoluminescence emission (PL) and excitation (PLE) spectra of the synthesized SrS with variable concentrations of $Ce^{3+}$ were investigated at room temperature (RT) to characterize the photophysical properties of our products, Fig. 3. In Fig. 3a, the SrS:$(Ce^{3+})_x$ phosphors exhibit two overlapping signals at λ≈495 nm and a minor shoulder centered at λ≈530 nm under excitation at λ≈276 nm. It is due to the typical $Ce^{3+}$ emission from the $5d^1$ excited energy level to the 4f$^1$ ground energy level.[26,34] The corresponding PLE spectrum (Fig. 3d) emitted at λ≈536 nm also possesses a broad band in the region of λ≈250–500 nm, accompanied by a weak band at λ≈280 nm and an intense band at λ≈430 nm. The former band (280 nm) can be sttributed to the effective host band-band transition whereas the latter (430 nm) is related to the transitions to the $^2T_{2g}$ and $^2E_g$ states of the 5d electronic manifold of octahedrally coordinated Ce$^{3+}$ replacing the Sr$^{2+}$ ion in the SrS lattice.[34,35] We conclude that SrS doped with Ce$^{3+}$ is sufficiently excited by both ultraviolet and blue light and shows its potential for LED fabrication.

Meanwhile, when doping a certain amount of the Ce, the PL peak maximum exhibits a slight red shift from 491 nm to 500 nm and the PL peak amplitude significantly decreases (Fig. 3c and d respectively). These effects are associated to an increased defect density in the crystal lattice and the migration of Ce$^{3+}$ ions during the crystallization process. In stoichiometric SrS, the substitution of Ce$^{3+}$ at the Sr$^{2+}$ site (Ce$_{Sr}$) is believed to be compensated by a Sr$^{2+}$ deficiency to form a charge-neutral Ce$_{Sr}$–V$_{Sr}$ complex. With increasing the Ce$^{3+}$ concentration, the Sr$^{2+}$ deficiencies into SrS:Ce$^{3+}$ were insufficient to fully compensate the excess negative charge of Ce$_{Sr}$, leading the decreased emission intensity.[36,37] Notably, a long afterglow is observed by the naked eyes and lasts about several seconds in both single Ce$^{3+}$ doped and double doped phosphors (Fig. 3a and Fig. S9), which further confirms the presence of Sr$^{2+}$ vacancy.

Within the doping range of the study, 0.5 mol% Ce$^{3+}$ is considered the optimized doping level realized with the employed solid-state method. Hence it is assumed that the likelihood of energy transfer among the Ce$^{3+}$ ions increases with increasing concentration of Ce$^{3+}$ ions in the host matrix.[11] Judging by G. Blasse's proposal (1), the critical distance of energy transfer ($Rc$) is calculated as 13.75 Å.

$$Rc = 2\left(\frac{3V}{4\pi x_c N}\right)^{1/3} \tag{1}$$

where $V$ is the volume of the unit cell, $N$ is the number of total $Ce^{3+}$ sites per unit cell and $x_c$ is the critical concentration of the activator ion. In SrS, $V$=217.73 Å$^3$, $N$=4, and the critical concentration, $x_c$, is about 0.005 in our system. As the value of $Rc$ is over 5Å, the exchange interaction has no accountable effect for non-radiative energy transfer processes between adjacent $Ce^{3+}$ ion in the matrix. Thus, the concentration quenching mechanism in $Ce^{3+}$ doped SrS phosphors may have happened due to multipole-multipole interactions which is responsible for the energy transfer of forbidden transitions.[11,38]

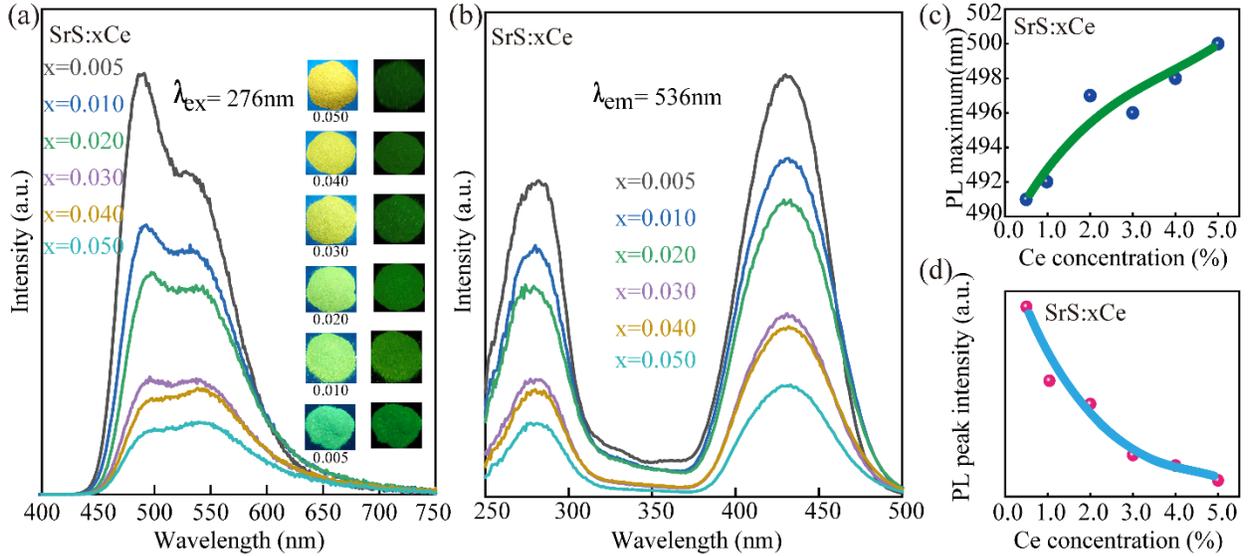

Figure 3. (a) PL spectra of SrS:$(Ce^{3+})_x$ and the inset are photos of samples under 365 nm UV lamp and with afterglow luminescence. (b) PLE spectra of SrS:$(Ce^{3+})_x$. (c,d) Dependence of PL maximum position and PL peak intensity on Ce concentration of SrS:$(Ce^{3+})_x$, respectively..

The Commission International de L'Eclairage (CIE) 1931 chromaticity coordinates calculated from the emission spectra of SrS:(Ce$^{3+}$)$_x$ phosphors are shown in Fig. S4. The emission color adjusted from green to green-yellow with increasing Ce$^{3+}$ concentration, suggesting increased crystal-field splitting of the Ce-5d energy levels by Ce$_{Sr}$ and V$_{Sr}$ substitution, while the corresponding CIE chromaticity coordinates ranged from (0.2465, 0.4798) to (0.3269, 0.5173).[39] Compared with other green phosphors, the value along y axis in the CIE chromaticity is larger than that of (Ba$_{1.2}$Ca$_{0.8-x}$Eu$_x$) SiO$_4$ and Ca$_2$YHf$_2$Al$_3$O$_{12}$:Ce$^{3+}$, Tb$^{3+}$. Thus, the as-prepared phosphors have greater luminescence and show more yellow color tune.[40,41] The CIE color coordinates and correlated color temperature (CCT) are listed in Table S1. These results indicate that the phosphors can exhibit tunable properties for various SS-LED applications and are highly promising candidates for general lighting devices.

In order to sustain electric neutrality for enhancing the luminescent properties, Na$^+$ alkali-metal ions were introduced into 1 mol% Ce$^{3+}$ based phosphors SrS. The concentration of Na$^+$ varies from y=0.5 to y=3.0 mol% by the aid of conventional solid-state method.[34] As expected in Fig. S5, the major reflections in our diffraction patterns can match well with the base SrS patterns, demonstrating that the introduction of Ce$^{3+}$ and Na$^+$ does not distort the long-range crystal structure of the host lattice. Similar with the data shown in Fig. 1, higher angle shifts are observed for the diffraction peaks between 2θ=50° and 53° degrees after the doping with Na$^+$ ions (Fig. S6). According to Bragg's law (2) and Scherrer's formula (3), we conclude that the structure change caused by Na$^+$ doping is negligible.

$$n\lambda = 2d\sin\theta \tag{2}$$

$$D = \frac{k\lambda}{\beta\cos\theta} \tag{3}$$

The representative FE-SEM images and EDS mapping results of $Ce^{3+}$ doped SrS with 0.5 mol% Na are displayed in Fig. 4. From the SEM images, it is demonstrated that the sample was composed of anomalous and aggregated particles with a size of 5-30 microns (Fig. 4a and 4b). To investigate the chemical composition of the particles, the EDS spectra confirms the uniform distribution of Sr, S, Ce and Na, suggesting successful doping of $Ce^{3+}$ and $Na^+$. Moreover, the elemental distribution was also investigated by EDS, which yields an average Sr/S atomic molar ratio of ~100% ((Fig. 4c-e and Fig. S7), in accordance with our XRD patterns. Ce and Na are homogenously distributed exhibiting low intensities in the target sample (Fig. 4f and 4g).

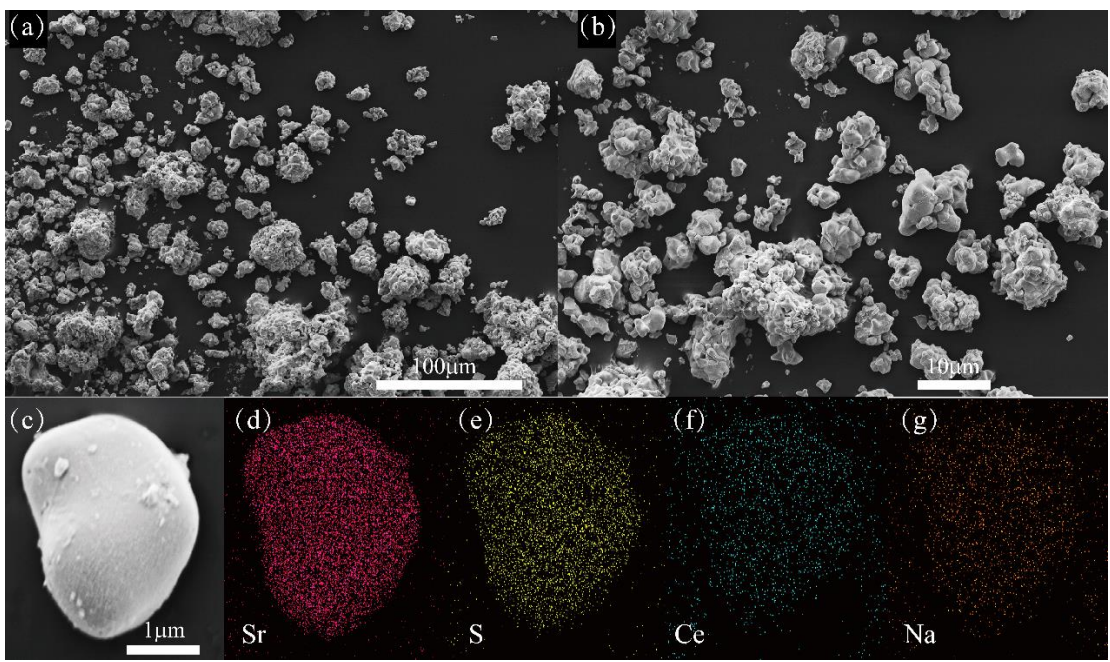

Figure 4. (a) and (b) SEM images of SrS with 1.0 mol% $Ce^{3+}$ and 0.5 mol% $Na^+$ dopant. (c) An enlarged particle for (d to g) element mapping images of Sr, S, Ce and Na in the selected particle.

It is widely acknowledged that the phosphor with efficient charge compensation shows enhanced luminescent intensity.[42-46] In addition, the ionic radius of charge compensator plays a vital role in the host lattice. In the case of SrS: $Ce^{3+}$, the excess of positive charge caused by $Ce^{3+}$ in the matrix were well compensated with $Na^+$ and the ionic radius of $Na^+$ (1.16Å) is close to that of $Sr^{2+}$ (1.18 Å). To further explore the effect of $Na^+$ in the $Ce^{3+}$-based material, the PL and PLE spectra were assessed. As given in Fig. 5a, all the PL spectra of $Na^+$ doped $SrS:(Ce^{3+})_{0.01}$ powders show broad-band extending within the wavelengths λ=400-700 nm under 276 nm UV light excitation. In Fig. 5d, monitored at 486 nm, the PLE spectra of $SrS:(Ce^{3+})_{0.01},(Na^+)_y$ samples show a signal with weak amplitude at λ≈430 nm and a strong signal with large amplitude at λ≈276 nm, which is possibly associated to the 4f→5d transitions of $Ce^{3+}$ in an octahedral crystal field. Fig. S8 shows a comparison of obtained PL spectra of $SrS:(Ce^{3+})_{0.01},(Na^+)_x$, (0.5% ≤ x ≤ 3%). Comparison of the spectra, photoluminescence is almost independent of $Na^+$ doping, despite minor fluctuations in signal intensities. PL peak amplitudes were found to increase with $Na^+$ concentration of [0, 1.5] mol% and decreases upon further increasing $Na^+$ concentration, Fig. 5b and c. Significantly, with increasing $Na^+$, the PL intensity is enhanced and the peak is blue shifted up to the maximum of 1.5 mol% $Na^+$ doping. In a previous report, Tong et al. found the incorporation of the charge compensator onto the Sr lattice site of $SrS:Ce^{3+}$ minimizes vacancy formation, which can produce a blue shift in the PL spectra.[47] In the same vein, as $Na^+-S^{2-}$ bond lengths in SrS are smaller than that in $Sr^{2+}-S^{2-}$ bond lengths, this site-perturbation should be more pronounced when the concentration of $Na^+$ is higher than that of of pristine samples. Thus, the red shift of the emission spectrum occurs in samples with higher doping concentration of $Na^+$, as pointed out previously.[48] Based on the results, doping mechanism and possible electronic transitions are summarized in Fig. 5e. As seen in Fig. S9, there is no significant difference of the emission color, which changes only from cyan to green as $Na^+$ ion is gradually increased. The CIE color coordinates and CCT were listed in Table S2. To demonstrate the luminous effect of the as-prepared phosphors for potential practical application, SrS:(Ce$^{3+}$)$_{0.01}$,(Na$^+$)$_{0.015}$ was incorporated in a w-LED lamp. As a specific package process shown in Supporting information, a commercial blue InGaN chip ($\lambda_{em}$ = 430 nm) was used for packaging experiment.

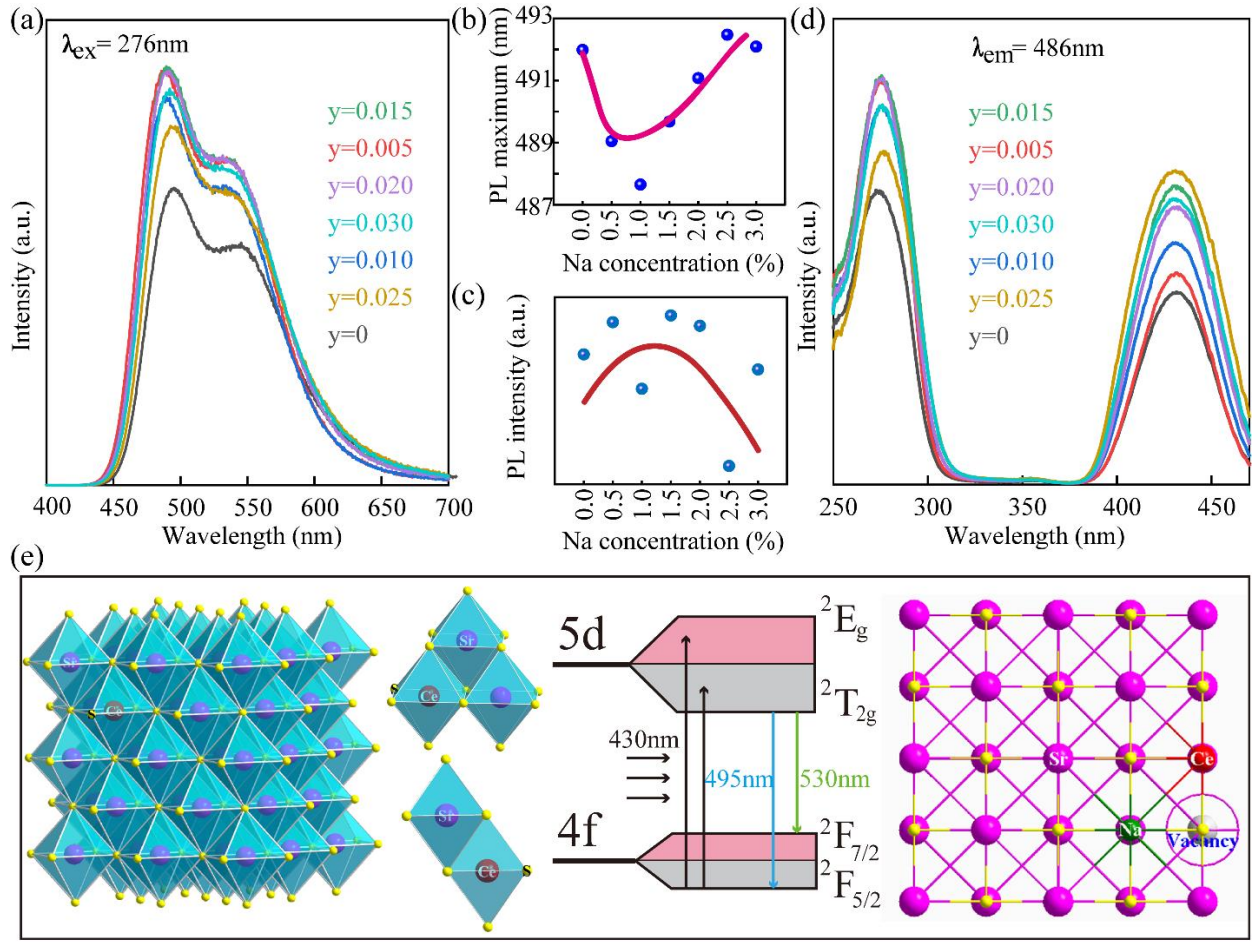

Figure 5. (a) PL spectra of SrS:1mol%Ce$^{3+}$,yNa$^+$. (b) and (c): dependence of PL maximum position and peak intensity on Na concentration of SrS:1mol%Ce$^{3+}$,yNa$^+$, respectively. (d) PLE spectra of SrS:1mol%Ce$^{3+}$,yNa$^+$.

Fig.6 shows the emission spectrum of packaged lamp driven by a 99.95 mA current at 2.821 V, which consists of the green and yellow light from the as-prepared phosphors. The red light was compensated by nitride phosphor and the Ra shifted from 60 to 89.7. As shown in Fig. S9, the corresponding CIE color coordinates of the packaged w-LEDs lamp are calculated to be (0.3108, 0.3458), which are similar to ideal white chromaticity coordinate (0.3333, 0.3333) of the National Television Standard Committee system. The inset photo in Fig.6 displays an intense white light emitting for the packaged lamp with CCT = 4823 K and high CRI value of 87.0. Moreover, the afterglow phosphor shows large flexibility in alternating current (AC) and can completely eliminate the use of an AC/ direct current (DC) converter required in conventional LED lighting technologies, thereby leading to reduced cost and further enhanced efficiency.[49,50]

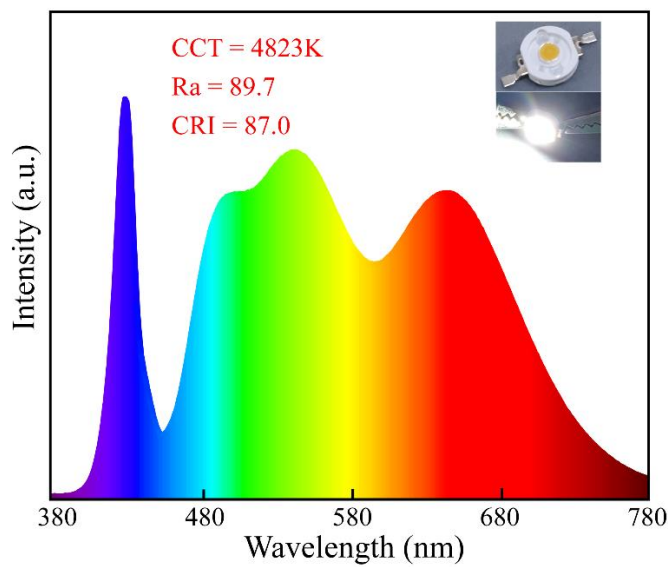

Figure 6. PL spectrum and photos of the w-LED device fabricated with as-prepared phosphors, the commercial red phosphor $Sr_2Si_5N_8:Eu^{2+}$ and an LED chips (λ=430 nm).

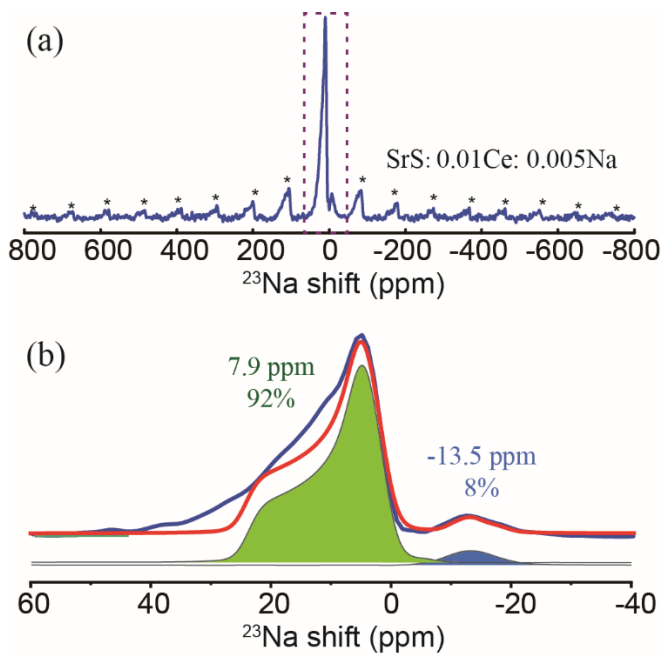

Figure 7. (a) Solid-state $^{23}$Na-NMR spectrum of SrS:(Ce$^{3+}$)$_{0.01}$,(Na$^{+}$)$_{0.005}$. Asterisks indicate spinning side bands (SSBs). (b) Zoomed in spectral region around the centroid of the spectrum. The experimental data (blue line) was simulated using the algorithm of Hughes and Harris et al.[51] The shift position and relative occupancy are noted next to the corresponding Na assignments.

The static $^{23}$Na-NMR spectrum (envelope over the MAS spectrum shown in Fig. 7a) displays a broad signal of 1600 ppm (about 180 kHz). Spectral simulation of the centroid region of the MAS spectrum (at $(\omega - \omega_0) = \pm 60$ ppm) reveals one nearly axially symmetric coordination site of sodium ions with an isotropic frequency shift of $(\omega - \omega_0)_{iso} = 7.9(2)$ ppm and a pronounced anisotropy of 10.5(3) ppm. Additionally, we observe a second very weak signal at about $(\omega - \omega_0)_{iso} = -13.5(8)$ ppm contributing only up to 8% of the total signal intensity and may be originating in surface ions. Further studies will be conducted to illuminate this possibility.

The poor fit quality of the main peak may be caused by a superposition of both angular dependent chemical and paramagnetic shift interactions influencing the resonance frequency distribution of non-coaligned principal axis systems.[51,52] The NMR results indicate that Na is successfully doped into the lattice of SrS, rather than the formation of $Na_2S$.[19]

We have successfully designed and synthesized SrS:($Ce^{3+}$),($Na^+$) phosphor exhibiting a broad emission band in the wavelength range of λ=430–700 nm via a facile solid state method. By designing Na-substituted and Sr-deficient SrS:($Ce^{3+}$),($Na^+$) to realize charge-compensating defects, higher emission intensity and afterglow phenomena were generated. The luminescence properties were modulated by modification of the crystal structure via doping with rare-earth metal ions and thus influencing crystal-field splitting at the octahedral coordination site of the Cerium ions.

The well-packaged w-LED lamp exhibits a warm white light (4823K) and a high CRI (Ra) value of 87.0 (89.7), promising the application of SrS:($Ce^{3+}$) phosphors as blue-excited green-emitting components for w-LEDs. The present demonstration of design strategy and characterization methods can also initiate further exploration to design various advanced phosphors with tunable spectra and improved performance for w-LEDs.

**Acknowledgement**

This work was financially supported by the National Natural Science Foundation of China (Grants 21974007, U1930401 and U1530402). The authors kindly thank Dr. Kuo Li for his insightful discussions and also help from Shijing Zhao, Wenfeng Peng, Wenbo Cheng, Jingbo Nan and Mingxing Chen for PL & PLE data acquisition.

**Conflict of interest**

The authors declare that they have no conflict of interest.

**Supporting Information**

The supporting information is available online or on request from the authors.